\documentclass[aps,prd,showpacs,preprintnumbers,nofootinbib,twocolumn]{revtex4}
\usepackage{dcolumn}
\usepackage{lipsum}
\usepackage{graphicx,epsfig,float}
\usepackage{psfrag}
\usepackage{bm}
\usepackage{epstopdf}
\usepackage{latexsym}
\usepackage{multirow}
\usepackage{amsmath,amssymb,amsfonts}
\usepackage{enumerate}

\def\beq {\begin{equation}}
\def\eeq {\end{equation}}
\def\bea {\begin{eqnarray}}
\def\eea {\end{eqnarray}}
\def\br{\begin{eqnarray}}
\def\er{\end{eqnarray}}
\def\nn {\nonumber}
\def\bc {\begin{center}}
\def\ec {\end{center}}
\def\l{\left}
\def\r{\right}
\def\dis{\displaystyle}
\def\thi{\thinspace}
\newcommand{\eel}[1] {\label{#1}\end{equation}}

\def\b  {\beta}
\def\c  {\gamma}

\newcommand{\bdm}{\begin{displaymath}}
\newcommand{\edm}{\end{displaymath}}

\def\p  {\pi}

\def\pa {\partial}

\def\bi {\begin{itemize}}
\def\ei {\end{itemize}}

\def\bc {\begin{center}}
\def\ec {\end{center}}

\DeclareMathOperator{\csch}{csch}

\begin{document}

\title{Entanglement entropy for nonzero genus topologies}
\author{S.\ Santhosh Kumar$^1$} \email[email: ]{santhu@iisertvm.ac.in}
\author{\ Suman Ghosh$^2$} \email[email: ]{suman.ghosh@boson.bose.res.in}
\author{S.\ Shankaranarayanan$^1$\footnote{Corresponding author}} \email[email: ]{shanki@iisertvm.ac.in}

\affiliation{$^1$\noindent School of Physics, Indian Institute of Science Education and Research(IISER-TVM), Thiruvananthapuram- 695 016, India}
\affiliation{$^2$\noindent Department of Theoretical Sciences, S.N. Bose National Centre for Basic Sciences, Salt Lake, Kolkata- 700 098, India}

\begin{abstract}
Over the last three decades entanglement entropy has been obtained for
quantum fields propagating in Genus-0 topologies (spheres).  For
scalar fields propagating in these topologies, it has been shown that
the entanglement entropy scales as area. In the last few years
nontrivial topologies are increasingly relevant for different
areas. For instance, in describing quantum phases, it has been
realized that long-range entangled states are described by topological
order. If quantum entanglement can plausibly provide explanation for
these, it is then imperative to obtain entanglement entropy in these
topologies.  In this work, using two different methods, we explicitly
show that the entanglement entropy scales as area of the Genus-1
geometry.
\end{abstract}


\maketitle

\section{Introduction}

Black holes have posed many puzzles, such as the  information paradox and
origin of Bekenstein-Hawking entropy, which are related to fundamental
principles of general relativity and quantum physics. Although there
have been several proposals to understand these, none of
these have provided any consistent framework \cite{EE-BH} (For recent
reviews, see Refs. \cite{wald,shanki-sour,solod}). Quantum entanglement is
one approach that naturally provide physical understanding of some of
these puzzles \cite{shanki-sour,solod}. Specifically, for black holes,
entropy due to entanglement naturally refers to the measure of the
information loss (for an outside observer) due to the spatial
separation between the degrees of freedom inside and outside the
horizon.
 	
It is well known that the entanglement entropy (EE) or the so-called
geometric entropy follows the so-called area law, first demonstrated
by Bombelli {\it et\; al} \cite{1986-Bombelli.etal-PRD} and Srednicki
\cite{1993-Srednicki-PRL}.  The EE can been derived via many approaches,
e.g. in the context of conformal field theories using the so-called
replica method \cite{Holzhey-etal-1994, Callan-1994,
  Cardy-Calabrese-2004,Vidal-etal-2013,Latorre-etal-2004}.  This
method is also applied to compute the EE for horizons with conical
singularities as such \cite{Solodukhin, Casini-2006, Casini-2010}.  In
recent years the EE has been  found to play crucial roles in understanding
many quantum phenomena and their applications
\cite{2009-Horodecki.etal-RMP,2010-Eisert.etal-RMP}.  A holographic
definition of the EE \cite{Ryu-Takayanagi-prl-2006} is proposed as a
universal formula to compute entropy of a black hole, in any
dimension, using AdS/CFT correspondence, and further attempts are being
made to understand its implications
\cite{Ryu:2006ef,Casinietal-2011,LM-2013}.

Studies on higher-dimensional black holes have become very crucial in
order to differentiate the generic features of the black holes with
the dimensional specific features.  Attempts have been made to
investigate thermodynamic properties of higher-dimensional black holes
in string theory \cite{Cvetic-1996,Chamblin-1999} and loop quantum
gravity \cite{Norbert-2013}.  In Ref. \cite{Ma-2006}, using the  Euclidean path
integral approach, it was shown that higher-dimensional (spherically
symmetric) rotating black holes obey the Bekenstein-Hawking entropy
formula.

Recently, Emparan and Reall obtained an exact solution of a
five-dimensional black hole with an event horizon of topology $S^1
\times S^2$ \cite{BR-prl2002,BR-review-2006}. These objects can be
understood as a circular neutral black string in five dimensions,
constructed as the direct product of the Schwarzschild solution and a
circle. However, the string has to rotate along $S^1$ to be
stable. The rotating black ring solutions have been rederived in a
systematic manner via solution-generating techniques in Refs.
\cite{solitons-2006,dipoles-2006}.  Note that these classes of black
holes are not only examples of nonspherical horizon topology but are
counterexamples to black hole uniqueness; i.e. the  no-hair theorem does
not extend to higher dimensions. 

Following the work of Emparan and
Reall, there have been studies to understand the thermodynamic
properties of a black ring. Exact microscopic entropy of
nonsupersymmetric extremal black rings is exactly reproduced for all
values of the ring radius using the same conformal field theory of the
four-charge four-dimensional black hole in Ref. \cite{Emparan-cqg2008}.
For supersymmetric black rings \cite{Cai-Pang-2007}, entropy function
was found (from both on-shell and off-shell perspectives) to be
reproducing the Bekenstein-Hawking entropy.  Further, the higher-order
corrections to the entropy arising from the  five-dimensional Gauss-Bonnet
term and supersymmetric $R^{2}$ completion was also computed.  In
Ref. \cite{Larsen-2005}, a string theory description of near-extremal black
rings was proposed and the thermodynamic properties were derived for a
large family of black rings.  Earlier, in
Refs. \cite{Strominger-2005,Datta:2013hba} M theory was used to give an exact
microscopic accounting of the black ring entropy.

In this work, we compute the  EE for scalar field systems propagating with
 Genus-1 topology to investigate the robustness of the area law.  In
general we analyze entropy across entangling surfaces with
$S^{m}\times S^{n}$ topology. We explicitly compute the EE for $S^1
\times S^1$ and $S^1 \times S^2$ horizons.

Although the approaches in Refs. \cite{Holzhey-etal-1994, Callan-1994,
  Cardy-Calabrese-2004,Vidal-etal-2013,Latorre-etal-2004,Solodukhin,
  Casini-2006,
  Casini-2010,2009-Horodecki.etal-RMP,2010-Eisert.etal-RMP,Ryu-Takayanagi-prl-2006,Ryu:2006ef,Casinietal-2011,LM-2013}
provide analytic expressions for the EE, one needs to assume certain
symmetries for the underlying theories. However, Genus-1 topology has
lesser symmetry than the Genus-0 topologies and hence one cannot
use these approaches. In this work, we use {\it ab initio
  calculations} or the real-time approach used in
Refs. \cite{1993-Srednicki-PRL,Plenio:2004he,2008-Das.etal-PRD} to
compute the EE in Genus-1 topologies. The real-time approach has the
advantage over the approaches as one can test the robustness of the
area law in nonvacuum states \cite{2008-Das.etal-PRD} and arbitrary
dimensions with spherical horizons \cite{Braunstein:2011sx}.

As a warmup toward computing EE of massless scalar field propagating in the 
black rings, in sec. II, we develop the methodology in a $3+1$- dimensional toroidal coordinate
system (i.e. for an entangling surface of topology $S^1 \times S^1$).
It is well known that
the Helmholtz equation is not separable in toroidal coordinates
\cite{Janaki-Dasgupta-1990,Morse,spencer}.  However, to compute the EE one needs to perform
integration over angular dimensions.  This can be done in the case of thin
rings \cite{Kunz-review-2013} using perturbative expansion.  We
use three such approximate methods to simplify our computation and
compare the final outcomes for a consistency check.  Thereafter, in
Sec. III we apply the same technique to compute the EE of black rings
($S^1 \times S^2$).  Finally, we end with a brief discussion on the
implications of the results obtained and limitations of the
methodology developed here.

In this work, we use the  $(+,-,-,-)$ metric signature and set $\hbar = c =
1$. Numerical computations are done in MATLAB (R2010b and R2012A) for
the lattice size $N = 100$, and error in the evaluation of the
entanglement entropy is $10^{-5}$.

\vspace{.4cm}
 \section{Warm up: Entanglement entropy in Torus} 

\subsection{Toroidal coordinate system ($r,\phi_{_1},\phi_{_2}$)}
\label{sec:TorusGeo}

A Toroidal coordinate system in three-dimensional space is well known. For
detailed discussion on the coordinate systems, we refer the readers to
Refs. \cite{Morse,spencer}. In this subsection, we discuss a few salient
features of the coordinate that we use to compute to the EE in this
geometry.

\begin{figure}[h]
\includegraphics[width=0.95\columnwidth]{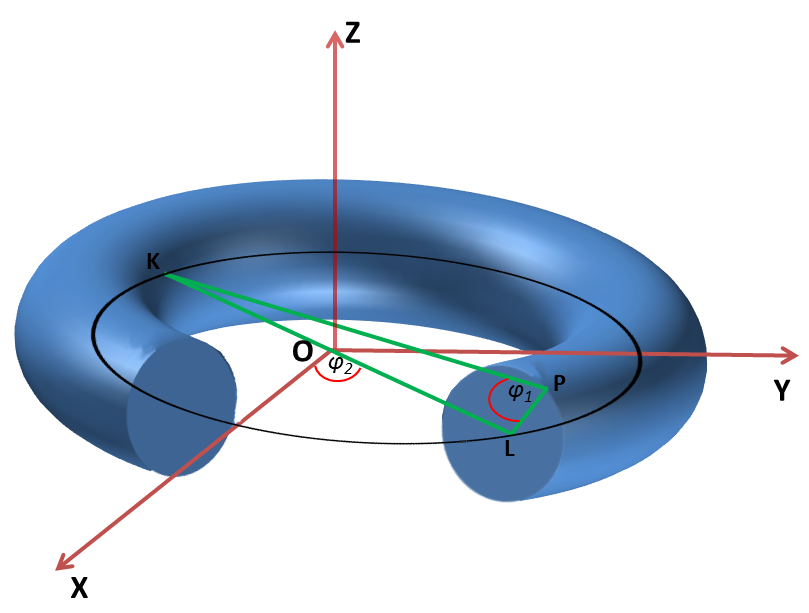} 
\caption{\footnotesize{Toroidal coordinate system($r,\phi_{_1},\phi_{_2}$)}}
\label{torus}
\end{figure}

It is an orthogonal coordinate system that results from rotating the
two-dimensional bipolar coordinate system by an angle $\phi_{_2}$
about the $Z$ axis [See Fig. (\ref{torus})].  The two foci are
separated by the focal line $KL$. The focal line lies in the $X-Y$
plane and is of length $2q$. It passes through the origin $O$ and
makes an angle $\phi_{_2}$ with $X$ axis.  The location of any point
$P$ in this space is given by ($r,\phi_{_1},\phi_{_2}$), where
$$ r=\ln \left(\frac{PL}{PK}\right), ~~ \angle KPL=\phi_{_1}$$   

The transformation relation between the rectangular ($X,Y,Z$) and the
toroidal coordinates is given by

\beq
\label{transfor}
 (X,Y,Z)=\frac{q}{\Delta}( \sinh r\cos\phi_{_2},\sinh r\sin\phi_{_2},\sin\phi_{_1})
\eeq
where, $\Delta=\cosh r -\cos\phi_{_1}$. The value range of the coordinates is  
$0\leq r < \infty$, $-\pi\leq \phi_{_1}\leq \pi$, $0\leq\phi_{_2}\leq 2\pi$.

The torus generated for any surface of constant $r$ is given by,
$$ Z^2+\left(\sqrt{X^2+Y^2}-q \coth r\right)^2 =\l(q \csch r\r)^2 $$

\noindent with the center at ($0,q \coth r $) in the $Z-Y$ plane. The
outer radius ($ R=q \coth r $) and inner radius ($\rho= q \csch r$) of
the torus are related to its focal length viz., $R^2-\rho^2 = q^2 $ and
the metric in this $3+1$-dimensional space -time is

\beq
\label{torousmetric}
ds^2=dt^2-\left(\frac{q}{\Delta}\right)^2 \left(dr^2+d\phi_{_1}^2+\sinh^2 r\thinspace d\phi_{_2}^2\right)
\eeq

\subsection{Approaches to compute entanglement entropy in toroidal geometry}

The action for the massless scalar field $\hat\Phi$ propagating in the
above background is
\beq
\label{torousaction}
S=\frac{1}{2}\int dt \thi d^3{\bf r} \sqrt{-g} \thi g^{\mu\nu} \partial_\mu\hat\Phi\partial_\nu \hat\Phi
\eeq
The form of the action  is,
\bea
\label{tor_acti_form}
S&=\dis\frac{1}{2}\int dt \thi d^3{\bf r}\frac{q^3}{\Delta^3}\sinh r\l[(\partial_t\Phi)^2-\l(\frac{\Delta}{q}\r)^2 \dis\l[ \l(\partial_r\Phi\r)^2\r.\r.\nn\\
&\l.\l.\dis+(\partial_{\phi_{_1}}\Phi)^2+\frac{1}{\sinh^2r}(\partial_{\phi_{_2}}\Phi)^2 \r]\r].
\eea
As mentioned earlier, the Helmholtz equation is not separable in
toroidal coordinates \cite{Morse,spencer}.  This implies that the
scalar field wave functional cannot be decoupled and the Hamiltonian
of the field cannot be written as a product of functionals that
depend on only one variable.  In the rest of the section we use two
approximate --- {\it perturbative} and {\it constant angle} ---
approaches to evaluate the EE.

One can use two different schemes for the perturbative approach. The
first scheme is to assume that the inner radius is much smaller than
the outer radius. In the leading order this scheme leads to an action
similar to that of the scalar field in Genus-0 topology. In Appendix
\ref{app:approx1}, we discuss this approximation and show that the
entropy-area relation is satisfied. The second scheme, which is
discussed in the rest of this section, is to consider the limit in
which the inner radius $(\rho)$ is much smaller than the focal line
KL. Unlike the earlier scheme, at all orders of approximation,
Genus-1 topology effects will be retained.  Under this assumption, we
perturbatively expand the action (\ref{tor_acti_form}) in terms of the
inner radius of the torus, in terms of the the dimensionless parameter
$x = \rho/q$.

In the constant angle approach, we fix one of the angles of the
Genus-1 topology and evaluate the entropy for the scalar field. The
advantage of the constant angle approach compared to the perturbative
approach is that the EE can be computed exactly without any
approximation. In Appendix \ref{app:constantGenus0}, we show that
constant angle approach gives entropy-area relation for all dimensions
greater than 2 for Genus-0 topology. We also discuss the importance of
this approach. The constant angle $\phi_{_1}$ approach for the torus
geometry leads to sphere of radius $q/\sin\phi_{_1}$, centered at
$(0,0,q\coth\phi_{_1})$, and most part of its entangling surface is
outside the torus\cite{spencer,arfken} .  However, for the black
rings, this approach retains Genus-1 topology. We discuss this more
in the next section.

\subsection{Perturbative approach} 

We use the following ansatz to expand the scalar field in the toroidal geometry (\ref{torousmetric})
\beq
\label{ansatz1}
\hat\Phi( {\bf r},t)= \dis \sum_{\substack{ m_{_1},m_{_2}\\=-\infty}}^\infty 
 \frac{\hat\Psi_{m_{_1},m_{_2}}(r,t)}{\pi}\cos m_{_1} \phi_{_1} \cos m_{_2} \phi_{_2}
\eeq
As mentioned earlier, we expand the action in terms of the
dimensionless parameter $x=\rho/q$. The form of the action up to the
first order in $x$ is
\bea
 \label{pertu_action1}
S&\simeq\dis\frac{1}{2}\int dt d^3{\bf x}\thinspace  q^3 \l\{\l(\partial_t{\Psi}\r)^2 -\frac{1}{q^2x^2}\l[\l(\partial_{\Phi_{_1}}\Psi\r)^2\r.\r.\nn\\
 &\l.\l.\dis +x^3\l[\partial_x\l(\frac{\Psi}{\sqrt{x}}\r)\r]^2+x^2\l(\partial_{\Phi_{_2}}\Psi\r)^2\r]\r\}\nn\\
&\dis+\frac{1}{2}\int dt d^3{\bf x}\thinspace  q\l[-\frac{\cos\phi_{_1} \Psi^2}{x}+ 2 x \cos \phi_{_1} \l(\partial_x \Psi\r)^2 \r.\nn\\
&\l.\dis +\cos \phi_{_1} \Psi\partial_x \Psi +\frac{2}{x}\cos \phi_{_1}\l(\partial_{\phi_{_1}} \Psi\r)^2 \r.\nn\\
&\l.\dis -\frac{3}{x}\sin \phi_{_1}\Psi\partial_{\phi_{_1}} \Psi   +2 x \cos \phi_{_1} \l(\partial_{\phi_{_2} }\Psi\r)^2 \r]
\eea
where 
$$\Psi({\bf x},t)=\displaystyle\frac{\Phi({\bf
    x},t)\sqrt{x}}{(1+x^2)^{1/4} \l(\sqrt{1+x^2}-x
  \cos\phi_{_1}\r)^{3/2}} \, .$$

Substituting Eq. (\ref{ansatz1}) in Eq. (\ref{pertu_action1}) gives 
\bea
  \label{pertu_action2}
S&=\dis\frac{1}{2}\sum_{m_{_1},m_{_2}}^\infty \int dt \thinspace d\rho  \l[\l(\partial_t{\widetilde{\Psi}}_{m_{_1},m_{_2}}\r)^2 
-\l[\frac{m_{_1}^2}{\rho^2}\widetilde{\Psi}^2_{m_{_1},m_{_2}}\r.\r.\nn\\
&\l.\l.\dis +\rho\l[\partial_\rho\l(\frac{\widetilde{\Psi}_{m_{_1},m_{_2}}}{\sqrt{\rho}}\r)\r]^2+ \frac{m_{_2}^2}{q^2}\widetilde{\Psi}^2_{m_{_1},m_{_2}}\r]\r.\nn\\
&\l.\dis +\frac{2\rho}{q} \l(\partial_\rho\widetilde{\Psi}_{m_{_1},m_{_2}}\r)^2 
+\dis\frac{1}{2q}\l(\partial_\rho\widetilde{\Psi}^2_{m_{_1},m_{_2}}\r)\r.\nn\\
&\l.\dis +\l(\frac{2\thinspace m_{_1}^2-1}{\rho q}+\frac{2 \rho\thinspace m_{_2}^2}{q^3}\r)\widetilde{\Psi}^2_{m_{_1},m_{_2}}\r]
\eea
where $ q \Psi _{m_{_1},m_{_2}} =\widetilde{\Psi} _{m_{_1},m_{_2}}$
and we assume that $ \Psi_{m_{_1}+1,m_{_2}}
=\Psi_{m_{_1}-1,m_{_2}}$.\footnote{The approximation is to ensure that
  the  angular modes are independent of each other.} The Hamiltonian
corresponding to the above action is given by
\bea
\label{pertur_with_out_discr}
H&=\dis\frac{1}{2}\sum_{m_{_1},m_{_2}}^\infty \int d\rho  \l[\widetilde{\Pi}_{m_{_1},m_{_2}}^2
 -\frac{2\rho}{q}\l(\partial_\rho\widetilde{\Psi}_{m_{_1},m_{_2}}\r)^2\r.\nn\\
&\l.\dis +\left(\frac{m_{_1}^2}{\rho^2}+\frac{m_{_2}^2}{q^2}
\dis +\frac{1-2 m_{_1}^2}{\rho q}-\frac{2 m_{_2}^2 \thinspace\rho}{q^3}\right)\widetilde{\Psi}^2_{m_{_1},m_{_2}}\r.\nn\\
&\l.+ \dis\rho\thinspace\left[ \partial_{\rho}\left(\frac{\widetilde{\Psi}_{m_{_1},m_{_2}}(\rho)}{\sqrt{\rho}}\right)\right]^2\dis-\frac{\widetilde{\Psi}_{m_{_1},m_{_2}}}{q}\partial_\rho\widetilde{\Psi}_{m_{_1},m_{_2}}\r]
\eea
The evaluation of the density matrix requires the discretization of
the Hamiltonian. Let $a$ be the discretization scale and
$\rho=j\thinspace a,1\leq j\leq N$ with constant outer radius, say
$Q$, such that $N<<Q$.  Using the midpoint discretization scheme i.e.,
the derivative of the form $f(x) d_x[g(x)]$ is replaced by $f_{j +
  1/2} [g_{j + 1} - g_j]/a$, the discretized Hamiltonian is given by
\bea
 \label{pertur_hamil}
H&=\dis\frac{1}{2 a}\sum_{m_{_1},m_{_2}}\sum_{j=1}^N\l[\widetilde{\Pi}^2_{m_{_1}m_{_2},j}+\l(\frac{m_{_1}^2}{j^2}+\frac{m_{_2}^2}{Q^2}\r.\r.\nn\\
&\l.\l.\dis +\frac{1-2 m_{_1}^2}{Q j} -\frac{2 m_{_2}^2\thinspace j}{Q^3}\r)\widetilde{\Psi}^2_{m_{_1}m_{_2},j}\r.\nn\\
&\l.\dis +\l(j+\frac{1}{2}\r)\l(\frac{\widetilde{\Psi}_{m_{_1}m_{_2},j+1}}{\sqrt{j+1}}-
 \frac{\widetilde{\Psi}_{m_{_1}m_{_2},j}}{\sqrt{j}}\right)^2\r.\nn\\
&\l.\dis -\frac{\dis 2(j+\frac{1}{2})}{Q}\l(\widetilde{\Psi}_{m_{_1}m_{_2},j+1}-\widetilde{\Psi}_{m_{_1}m_{_2},j}\r)^2\r.\nn\\
&\l.\dis-\frac{\widetilde{\Psi}_{m_{_1}m_{_2},j}}{Q} \l(\widetilde{\Psi}_{m_{_1}m_{_2},j+1}-\widetilde{\Psi}_{m_{_1}m_{_2},j}\r)\r] 
\eea
\noindent where $\widetilde{\Psi}_{m_{_1}m_{_2},N+1}=0$. The
commutation relation between the dimensionless field operators is
given by
\beq 
\label{torous-commu2}
\left[\hat{\widetilde{\Psi}}_{m_{_1}m_{_2},j},\hat{\widetilde{\Pi}}_{m'_{_1}m'_{_2},j'}\right] = i\delta_{jj'}  \delta_{m_{_1},m'_{_1}}\delta_{m_{_2},m'_{_2}}
\eeq
The Hamiltonian in Eq. (\ref{pertur_hamil}) is in the form of a system
of $N$ coupled quantum harmonic oscillators
(\ref{eq:discretizeHamiltonian}) and can be written as an $N\times N$
symmetric semidefinite matrix (\ref{eq:Torus-Kij-Pert}).

Since the Hamiltonian is quadratic (\ref{eq:discretizeHamiltonian}),
the ground-state wave function can be written as
\bea 
\label{eq:GS-wavefn}
{\tilde \Psi}(x_1, \dots, x_N) ~=~ \left(\frac{|\Omega|}{\p^N}\right)^{1/4} 
\exp \left[-~ \frac{x^T \cdot \Omega \cdot x} 2\right] \, .
\eea
The corresponding density matrix can be evaluated exactly as \cite{1993-Srednicki-PRL}:
\beq 
\label{eq:GS-den}
\rho(t; t') = \sqrt{\frac{|\Omega|}{\pi^{N-n} |A|}} ~\exp \left[- \frac{t^T \c t 
+ t'^T \gamma t'} 2 ~+~ t^T \beta t'\right] \, ,
\eeq 
where we have decomposed
\bea 
\label{eq:Omega}
\Omega  \sim K^{1/2} ~=~ \left( \begin{array}{ll} 
{A} & {B} \\
{B^T} & {C} 
\end{array} \right) \, 
\eea
and defined
\beq
\beta ~=~ \frac{B^T A^{-1} B} 2 ~;~~~ \gamma ~=~ C - \beta  \, .
\eeq
$A$ is an $n' \times n'$ symmetric matrix, $B$ is an $n' \times
(N-n')$ matrix, and $C, \b, \gamma$ are all $(N-n') \times (N-n')$ symmetric
matrices. The matrices $B$ and $\beta$ are nonzero only when
the Harmonic oscillator (HO)s are interacting. 

Performing a series of unitary transformations,
\bea 
\label{eq:diag}
&& V \c V^T = \c_D = \mbox{diag}~,~ {\bar\b} \equiv \c_D^{- 1/2} 
V \b V^T \c_D^{- 1/2}~,\nn\\ 
&& W {\bar\b} W^T = {\bar\b}_D = \mbox{diag}~,~  v \equiv 
W^T \c_D^{1/2} V ~, 
\eea
one can reduce $\rho (t; t')$ to a product of the reduced density
matrices $\rho_{_{(2-\rm HO)}} (t; t')$ for $(N-n')$ two coupled HOs with 
one oscillator traced over (i.e., $N=2, n'=1$) \cite{1993-Srednicki-PRL},
\bea
\label{eq:GS-den1}
\rho(t; t')&=& \prod_{i=1}^{N-n}  \rho_{_{(2 - \rm HO)}} (t; t') \\
\rho_{_{(2-\rm HO)}} (t; t')& =& \sqrt{\frac{|\Omega|}{\pi^{N-n'} |A|}} 
\exp \left[- \frac{v_i^2+v_i'^2}{2} + {\bar\beta}_i v_i v_i' \right] \nn
\eea
where $v_i \in v$ and ${\bar\beta}_i \in \bar\beta$. Correspondingly,
the total entanglement entropy is a sum of $(N-n')$ two-HO entropies
$S_i^{(2-\rm HO)}  ~ (i = 1, \dots, N-n')$ which are obtained using
the Von Neumann relation
\beq 
\label{2HO-ent}
S_i^{(2-\rm HO)} = - \ln[1-\xi_i] - \frac{\xi_i}{1 - \xi_i} \ln\xi_i \quad, 
\quad  \xi_i = \frac{{\bar \beta}_i}{1+ \sqrt{1 - {\bar \beta}_i^2}} \, .
\eeq
The total entropy for the full Hamiltonian $H = \dis\sum_{m_{_1}, m_{_2}} H_{m_{_1} m_{_2}}$, 
Eq.(\ref{eq:discretizeHamiltonian}), is therefore given by
\bea
\label{eq:GS-ent}
S (n',N)& =& \sum_{m_{_1}, m_{_2}} S_{m_{_1} m_{_2}} (n',N)  \\
S_{m_{_1},m_{_2}} (n',N) &=& - \ln[1-\xi_l] - \frac{\xi_l}{1 - \xi_l} \ln\xi_l \, .
\eea
\begin{figure}[h]
\includegraphics[width=0.8\columnwidth]{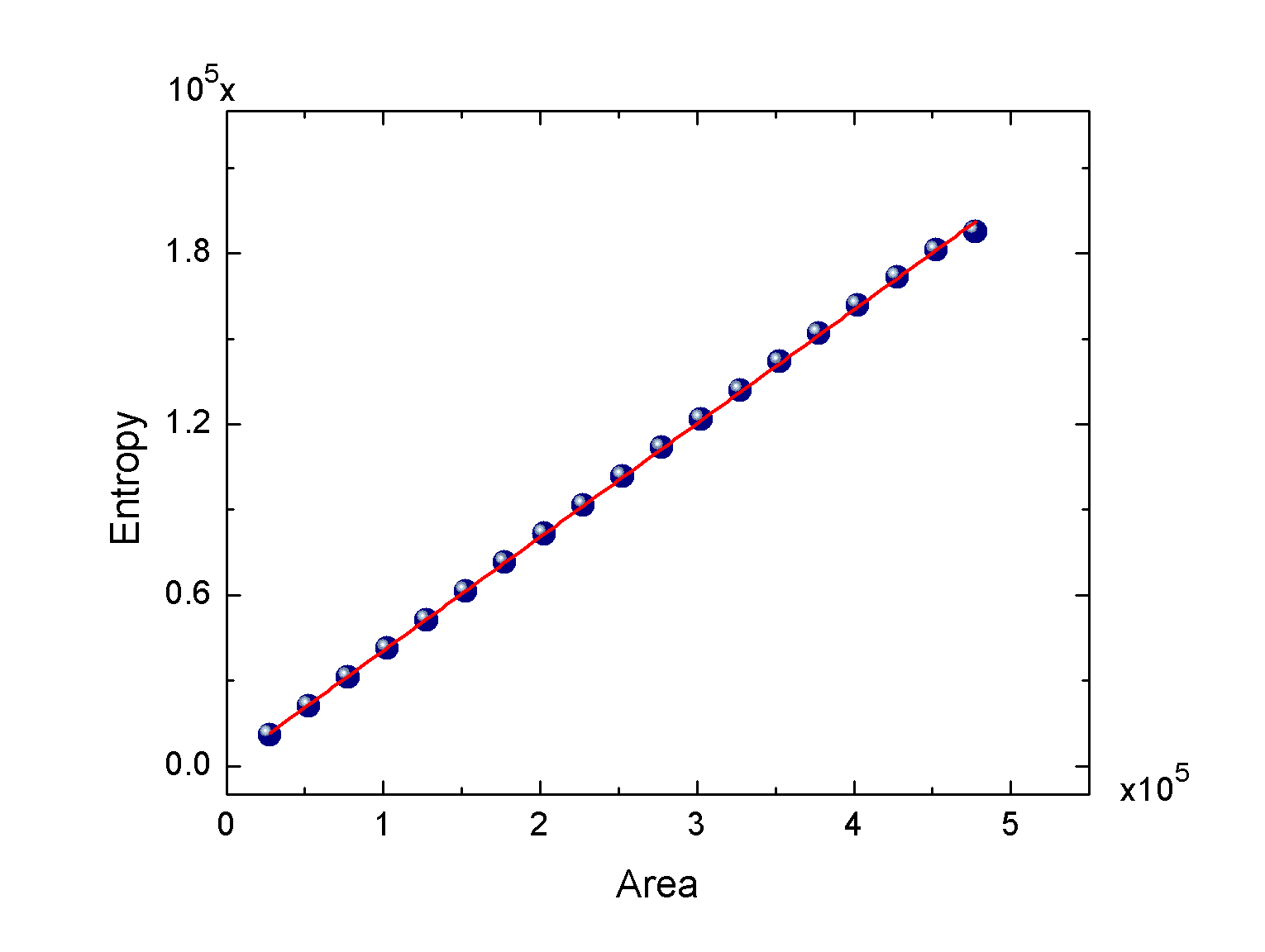} 
\caption{\footnotesize{Plot of entropy vs area-relation in the $3+1$ -dimensional torus using perturbative approach for $N = 100,~ Q=5000, ~5\le n'\le 90 
$. The blue dots are the numerical outputs and the red line is best linear fit.}}
\label{Fig:Torus-Pert}
\end{figure}
In Fig. \ref{Fig:Torus-Pert}, we have plotted entanglement entropy
vs area of the $S^1\times S^1$ surface.  As it is clear from the
figure, in the linear perturbative limit, the entropy is proportional
to area. We will discuss the importance of the result in
Sec. \ref{sec:Conclusion}.

\subsection{Constant angle approach}

As we mentioned earlier, the constant angle approach for the torus leads
to Genus-0 topology and does not provide any information about the
higher Genus topology. However, as discussed in the next section, it
does provide useful information in higher dimensions.  Below, we
discuss the procedure and obtain the EE for by setting $\phi_{_1}=\alpha$
for the torus ($\alpha$ is a constant).

Setting $\phi_{_1}=\alpha$ in the action (\ref{tor_acti_form}), the
Hamiltonian corresponding to the reduced action is,
\bea
\label{torous_cons_hami}
H&=\dis \frac{1}{2}\sum_{m} \int d\rho \l[\widetilde{\Pi}^2_{m} +\l(\frac{\rho\thinspace m\thinspace k_{_2}\thinspace\widetilde{\Psi}_{m}}{q^2}\r)^2\r.\nn\\
&\l. \dis +\frac{k_{_1}^2}{q^4}\l[ \partial_\rho \l( k_{_1}\rho\;k_{_2} \widetilde{\Psi}_{m}\r)\r]^2 \r]
\eea
where, 
\begin{align}
 k_{_1}(\rho)&= \dis\l(\rho^2+q^2\r)^{1/4}&k_{_2}(\rho)&=\dis\frac{k_{_1}^2}{\rho}-\cos\alpha\nn\\
\dis\dis\widetilde{\Psi}_{m}&=\frac{k_{_2}}{ q k_{_1}}  \Psi_{m}\nn
\end{align}

\begin{figure}[h]
\includegraphics[width=1\columnwidth]{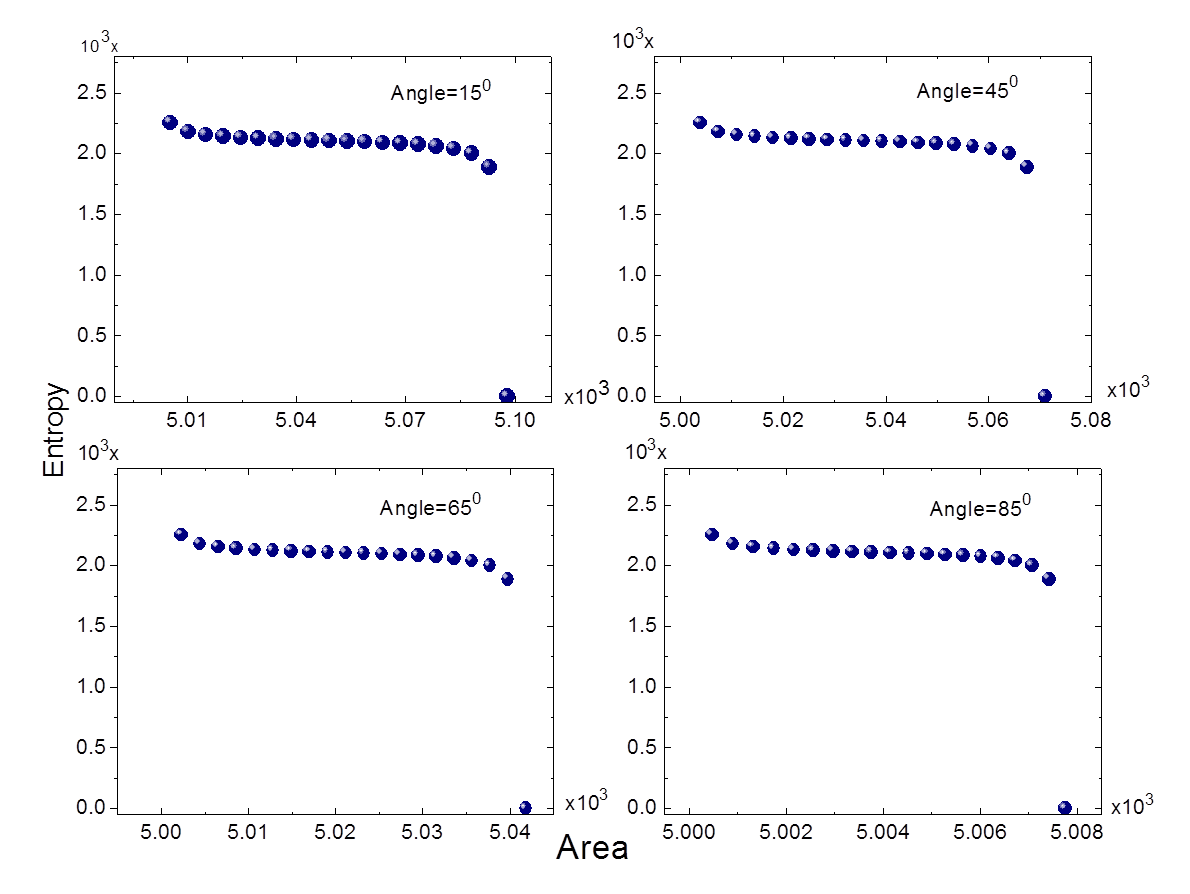} 
\caption{\footnotesize{Entropy vs scaled area profile in the $3+1$ -dimensional torus  using  {\it constant angle} approach for $N=100,~ Q=5000, ~5\le n'\le 90$. 
The blue dots are the numerical outputs.}}
\label{meth3(3)}
\end{figure}

As in the previous subsection, to evaluate the EE we need to discretize
the Hamiltonian. Here again, the lattice spacing is $a$ and the outer
radius is set to $Q$. Using the midpoint discretization scheme, the
above Hamiltonian is in the form of a system of $N$ coupled quantum
harmonic oscillators (\ref{eq:discretizeHamiltonian}) and can be
written as an $N\times N$ symmetric semidefinite matrix
(\ref{eq:Torus-Kij-ConstAngle}).

Following the procedure discussed in the previous subsection, we
obtain the EE for different angles.  Fig. \ref{meth3(3)} shows the
profile of the EE vs the larger radius for different constant angles. 
In the constant angle case, the entangling surface is 
a cylinder with flat sides and does not vary much with varying $\rho$ as is 
clear from Fig. \ref{meth3(3)}; i. e.,  the domain of the $x$ axis which is 
the scaled area of the entangling surface, decreases with increasing $\phi_1$,
resulting in a constant entropy. As we mentioned earlier, this approach for the torus does 
not provide insight on the entropy-area relation, but it helps to confirm 
our understanding of the nontrivial geometry of constant  $\phi_1$ surfaces. 

\section{Entanglement entropy in Ring Geometry} 

Ring geometry is a generalization of the torus for space-times greater
than $4$. In the case of five-dimensional space-time, the line element corresponding to
the ring coordinate ($S^{2} \times S^{1}$) is given in the recent
review of Emparan and Reall \cite{BR-review-2006}. In the following
subsection, we write down the line element for a general $S^{m} \times
S^{n}$, where $m,n$ are arbitrary integers with the restriction that
both of the them simultaneously cannot take the value $1$.  We then
focus on the specific case of evaluating the EE for $S^{2} \times S^{1}$
using the two approaches discussed in the previous section.

\subsection{Ring geometry}
\label{sec:RingGeo}
Let us consider the  $(m + n + 1)$ -dimensional line-element 
\beq
\label{eq:m+n+1Dspace}
ds^2 = \sum_{i=1}^{m} dx_i^2 + \sum_{j=1}^{n +1} dy_j^2  
\eeq
and perform the transformations
\begin{subequations}
\label{any_ring_coordinates}
\br
x_{_1}&=&r_{1} \cos\phi_{_1} \\
x_{_2}&=&r_{1} \sin\phi_{_1}\cos\phi_{_2} \\
\vdots &  & \vdots \nn \\
x_{_{m-1}}&=&r_{1} \sin\phi_{_1}\sin\phi_{_2}\ldots\cos\phi_{_{m-1}}\\
x_{_{m}}&=&r_{1} \sin\phi_{_1}\sin\phi_{_2}\ldots\sin\phi_{_{m-1}}\\
y_{_1}&=&r_{2} \cos\theta_{_1} \\
y_{_2}&=&r_{2} \sin\theta_{_1}\cos\theta_{_2} \\
\vdots & & \vdots \nn \\
y_{_{n}}&=&r_{2} \sin\theta_{_1}\sin\theta_{_2}\ldots\cos\theta_{_n}\\
y_{_{n+1}}&=&r_{2} \sin\theta_{_1}\sin\theta_{_2}\ldots\sin\theta_{_n}
\er
\end{subequations}
where $ 0\le r_{_1},r_{_2}<\infty$,\thinspace $0\leq\phi_{_1},\ldots,\phi_{_{m-2}},\theta_{_1},\ldots,\theta_{_{n-1}} \leq \pi,
\thinspace 0\le \phi_{_{m-1}},\theta_{_n}< 2\pi$. 
Substituting the above transformations in Eq. (\ref{eq:m+n+1Dspace}), we get
\bea
\label{general_line_el}
ds^2&=  dr_{_1}^2+r_{_1}^2 \l(d\phi_{_1}^2+\sin^2\phi_{_1}d\phi_{_1}^2+\ldots +\sin^2\phi_{_1}\ldots\r.\nn\\
&\l.\ldots\sin^2\phi_{_{m-2}}d\phi_{_{m-1}}^2 \r) +dr_{_2}^2+r_{_2}^2\l(d\theta_{_1}^2\r.\nn\\
&\l. +\sin^2\theta_{_1}d\theta_{_2}^2+\ldots +\sin^2\theta_{_1}\ldots\sin^2\theta_{_{n-1}}d\theta_{_n}^2\r)
\eea
The above line element corresponds to product to two spaces with the
symmetry $R \times S^{\alpha}$, where $\alpha$ is $m -1$ or $n$.
Performing the transformation
\beq
\label{ring_coord2}
r_{_1}=R \frac{\sin\theta}{\cos\theta+\dis\frac{R}{r}}\qquad r_{_2}=R \frac{\sqrt{\dis\frac{R^2}{r^2}-1}}{\cos\theta+\dis\frac{R}{r}}
\eeq
to the line element (\ref{general_line_el}), we get \cite{BR-review-2006},
\bea
\label{ring_general_metric}
ds^2&= \dis\frac{1}{(1+\dis\frac{r}{R}\cos\theta)^2}\l[\frac{dr^2}{1-\dis\frac{r^2}{R^2}}+r^2\l(d\theta^2 
+\sin^2\theta d\phi_{_1}^2\r.\r.\nn\\
&\l.\l.\dis +\ldots+\sin^2\theta\sin^2\phi_{_1}\ldots\sin^2\phi_{_{m-2}}d\phi_{_{m-1}}^2 \r)\r.\nn\\
&\l. +R^2(1-\dis\frac{r^2}{R^2})\l(d\theta_{_1}^2+\sin^2\theta_{_1} d\theta_{_2}^2+\r.\r.\nn\\
&\l.\l.\dis\ldots+\sin^2\theta_{_1}\ldots\sin^2\theta_{_{n-1}}d\theta_{_n}^2 \r)\r] \, .
\eea
Note that the ranges of $\theta$ and $r$ are $-\pi\leq\theta\leq
\pi$,\thinspace$0\leq r \leq R$.  It is also important to note that
$m$ and $n$ can take arbitrary values, with the restriction that both
of the them simultaneously cannot take the value $1$. For constant
$R$, the above line  element corresponds to most general Genus-1
topology $S^m\times S^n$. The Genus-1 topology is generated by the
surfaces of constant $r$:
\bea
x_{_1}^2+\ldots +x_{_m}^2+\dis \left( \sqrt{y_{_1}^2+\ldots +y_{_{n+1}}^2}-\dis \frac{R^2}{\sqrt{R^2-r^2}}\right)^2 \nn\\
 =\dis\frac{R^2 r^2}{R^2-r^2}
\eea
The simplest ring coordinates, as discussed in the review
\cite{BR-review-2006}, can be obtained by setting $m=2$ and $n=1$. For
$m=2$ and $n=1$, the line element (\ref{ring_general_metric}) reduces
to
\br
\label{ring_metric2}
ds^2&=&\dis\frac{1}{(1+\dis\frac{r}{R}\cos\theta)^2}\l[\frac{dr^2}{1-\dis\frac{r^2}{R^2}}+r^2\l(d\theta^2 +\sin^2\theta d\phi_{_1}^2\r)\r.\nn\\
&&\l.\dis +R^2\l(1-\frac{r^2}{R^2}\r)d\theta_{_1}^2\r]
\er
It is important to note that the ring geometry is such that the radius
($r$) of the 2-sphere cannot be larger than the radius ($R$) of the
circle.

In the case of five -dimensional space, one can embed two different
Genus-1 topologies, i. e., $S^2\times S^2$ and $S^3\times S^1$.  To
embed $S^3\times S^1$ topology in six -dimensional space, using the
 coordinate transformation on to the line element
(\ref{eq:m+n+1Dspace}),
\begin{align}
\label{ring_5d_coordnts}
x_{_1}&=r_{_1}\cos\phi_{_1} & x_{_2}&=r_{_1} \sin\phi_{_1}\cos\phi_{_2}\nn\\
 x_{_3}&=r_{_1} \sin\phi_{_1}\sin\phi_{_2}&y_{_1}&=r_{_2} \cos\theta_{_1}\;\;\; y_{_2}=r_{_2} \sin\theta_{_1}    
\end{align}
leads to the following line element:
\bea
\label{ring_5d_metric1}
ds^2&=\displaystyle\frac{1}{(1+\dis\frac{r}{R}\cos\theta)^2}\l[\frac{dr^2}{1-\dis\frac{r^2}{R^2}}+r^2\l(d\theta^2 +\sin^2\theta d\phi_{_1}^2\r.\r.\nn\\
&\l.\l.\dis +\sin^2\theta\sin^2\phi_{_1} d\phi_{_2}^2\r)+ \dis R^2(1-\frac{r^2}{R^2})d\theta_{_1}^2\r]
\eea
where $r_{_1}$ and $r_{_2}$ are given by Eqs.
(\ref{ring_coord2}). Here again the radius ($r$) of the 3-sphere cannot be larger than the radius ($R$) of the circle .

To embed $S^2\times S^2$ in six -dimensional space, one starts with
same transformation (\ref{ring_5d_coordnts}); however, changing the
definition for $r_{_1}$ and $r_{_2}$ in such way that
\beq
\label{ring_5d_radius}
r_{_1}=R \frac{\sqrt{\dis\frac{R^2}{r^2}-1}}{\cos\theta+\dis\frac{R}{r}}\qquad r_{_2}=R \frac{\sin\theta}{\cos\theta+\dis\frac{R}{r}}
\eeq
gives 
\bea
\label{ring_5d_metric2}
ds^2&=\dis\frac{1}{(1+\dis\frac{r}{R}\cos\theta)^2}\l[\frac{dr^2}{1-\dis\frac{r^2}{R^2}}+r^2\l(d\theta^2 +\sin^2\theta d\phi_{_1}^2\r)\r.\nn\\
&\l.\dis +R^2(1-\frac{r^2}{R^2})\l(d\theta_{_1}^2+\sin^2\theta_{_1} d\theta_{_2}^2\r)\r]
\eea
Here again, the radius ($r$) of the 2-sphere cannot be larger than
the radius ($R$) of the 2-sphere.

Similarly, we can embed $S^4\times S^1$ , $S^2\times S^3$ in
seven -dimensional space and $S^5\times S^1$, $S^4\times S^2$, $S^3\times
S^3$ in eight -dimensional space.
 
\subsection{Entanglement entropy of scalar fields in black ring}

Although the EE can be obtained for a general Genus-1 topology in
arbitrary dimensions, we focus on the specific case of black rings in
five -dimensional. This is mainly due to two reasons. First, as we go to
higher dimensions, it becomes numerically intensive. To compare the
computing time to calculate the entanglement entropy for one value for
black ring topology and the torus is 60 Peta Flop and 1 Peta Flop,
respectively. This increases exponentially as we go to higher
dimensions. Second, the torus is a special case of the Genus-1
topology. Comparing the transformations in Secs. \ref{sec:TorusGeo},
\ref{sec:RingGeo} it is clear that the transformations
(\ref{any_ring_coordinates}) cannot be used to obtain the torus
line element.  However, this is not the case for the higher -dimensional
generalization of Genus-1 topology.

\par The action in the $4+1$ -dimensional ring space-time is
\beq
\label{ring_action}
S=\frac{1}{2}\int dt \thinspace d^4{\bf r} \sqrt{g} \thinspace g^{\mu\nu} \partial_\mu\hat\Phi\partial_\nu \hat\Phi
\eeq
Using that the ring space-time is a product of $S^2$ and $S^1$, we use
the  ansatz for the scalar field:
\beq
\label{ring_ansatz}
\hat{\Phi}({\bf r},t)= \sum_{l=0}^\infty\sum_{m=-l}^l\sum_{n=-\infty}^\infty \frac{\chi_{_{l,m,n}}}{\sqrt{\pi}}\mathcal{Z}_{lm}(\theta,\phi_{_1}) \cos n\theta_{_1}
\eeq
where $\mathcal{Z}_{lm}$ is the real part of the spherical harmonics. 

As in the case of the toroidal background, it is not possible to separate
the Helmholtz equation in this background. Hence, it will not be
possible to define the ground state of this system exactly. Here
again, we use two different --- perturbative and constant angle ---
approaches to compute the entanglement entropy. The two approaches
provide complementary information about the entanglement entropy area
relation. The perturbative approach will be valid when  the radius
($r$) of the 2-sphere is much smaller than the radius ($R$) of the
circle. However, the constant angle approach is nonperturbative and
one can compute the EE for the case in which $r \simeq
R$. In the rest of this section, we calculate the EE for the scalar fields
using these two approaches.

\subsection{Perturbative approach}

Substituting the ansatz (\ref{ring_ansatz}) in the action
(\ref{ring_action}), perturbatively expanding the action up to the
first order in $r/R$ and integrating over the angular variables, we
get
\bea
 \label{ring_pert_actn}
S&=\dis\frac{1}{2}\sum_{lmn} \int \tilde{x}^2R^3 \thinspace d\eta d\tilde{x} \l[
 \l(\partial_\eta\Psi_{lmn}\r)^2-\l[\l(\partial_{\tilde{x}}\Psi_{lmn}\r)^2\r.\r.\nn\\
&\l.\l.\dis +\l(\frac{l(l+1)}{\tilde{x}^2}+n^2\r)\Psi^2_{lmn}\r]\r.\nonumber\\
&\l.\dis-\sum_{l'm'n'}   \l(\tilde{x}\partial_{\tilde{x}} \Psi_{lmn} \partial_{\tilde{x}} \Psi_{l'm'n'} I_{_1}
 +2  \Psi_{lmn} \partial_{\tilde{x}} \Psi_{l'm'n'} I_{_1}\r.\r.\nonumber\\
&\l.\l. \dis +\frac{1}{\tilde{x}} \Psi_{lmn}  \Psi_{l'm'n'} I_{_2} -\frac{2}{\tilde{x}} \Psi_{lmn}  \Psi_{l'm'n'} I_{_3}\r.\r.\nn\\
&\l.\l.\dis + \frac{1}{\tilde{x}} \Psi_{lmn}  \Psi_{l'm'n'} I_{_4}+n^2 \tilde{x}\Psi_{lmn}  \Psi_{l'm'n'} I_{_1} \r)\r]
\eea
where 
\beq
\Psi_{_{lmn}}({\bf x},t) =\displaystyle\frac{\chi_{_{lmn}}({\bf x},t)}{(1+\tilde{x}\cos\phi_{_1})^2} \, ,
\eeq
$\eta (= t/R)$ and $\tilde{x} (= r/R)$ are dimensionless variables;
and $I_{_1}$, $I_{_2}$, $I_{_3}$ and $I_{_4}$ are the integrals
involving the spherical harmonics given in Eq.
(\ref{pert_integrals1}). The solutions of these integrals are given in
Appendix D.
 
From the above action it is clear that lowest-order terms gives
nonzero values only when $l = l'$. This is consistent with the fact
the equation corresponding to the lowest-order action satisfies the 
Sturm-Liouville equation, and, by definition, each of these modes are
orthogonal to each other. However, the equation corresponding to the
full perturbed action cannot be written in the Sturm-Liouville equation,
and, hence it is natural that first order breaks this. Here, we set
$\Psi_{l+1,m,n}=\Psi_{l-1,m,n}$ which will capture the effect of the
first-order term.

Substituting Eq. (\ref{pert_integrals2}) in the above action and using
the above relation leads to
\bea
S&=\dis\frac{1}{2}\sum_{lmn} \int dt dr \l\{\l(\partial_t\widetilde{\Psi}_{lmn}\r)^2 -r^2\l[\partial_r\l(\frac{\widetilde{\Psi}_{lmn}}{r}\r)\r]^2\r.\nn\\
&\l.\dis -\l(\frac{l(l+1)}{r^2}+\frac{n^2}{R^2}+\frac{4C_0 l(l+1)}{rR}+\frac{4C_0 n^2 r}{R^3}\r)\widetilde{\Psi}^2_{lmn} \r.\nn\\
&\l.\dis - \frac{2 r C_0}{R} \pa_r\l(\frac{\widetilde{\Psi}_{lmn}}{r}\r) \l[ r^2\pa_r\l(\frac{\widetilde{\Psi}_{lmn}}{r}\r)+\widetilde{\Psi}_{lmn}\r] \r\}\nn
\eea
where 
\beq
\widetilde{\Psi}_{lmn}=r\sqrt{R}\Psi_{lmn} ,\quad C_0(l,m)=\dis\sqrt{\frac{l^2-m^2}{4l^2-1}}\sim C_1(l,m).
\eeq
Here again, it is important to note that the leading-order terms in
the action do not depend on the $m$ and have the degeneracy factor
$(2 l + 1)$; however, the first-order terms in the action (like $C_0$)
have explicit $m$ dependence. Although this does not have any physical
implications, it  has implications for the numerical
computations. Compared to the torus, here we need to evaluate the
entropy for each value of $m$ and hence the computation time
increases exponentially.
 
The Hamiltonian corresponding to the above action is
\bea
H&=\dis \frac{1}{2}\sum_{lmn} \int dr\l\{\widetilde{\Pi}^2_{lmn}+r^2\l[\partial_r\l(\frac{\widetilde{\Psi}_{lmn}}{r}\r)\r]^2\r.\nn\\
&\l.\dis +\l(\frac{l(l+1)}{r^2}+\frac{n^2}{R^2}+\frac{4C_0 l(l+1)}{rR}+\frac{4C_0 n^2 r}{R^3}\r)\widetilde{\Psi}^2_{lmn} \r.\nn\\
&\l.\dis + \frac{2 r C_0}{R} \pa_r\l(\frac{\widetilde{\Psi}_{lmn}}{r}\r) \l[ r^2\pa_r\l(\frac{\widetilde{\Psi}_{lmn}}{r}\r)
+\widetilde{\Psi}_{lmn} \r]\r\}
\eea
where $\hat\Pi_{lmn}(r)$ is canonically conjugate to $\hat\Psi_{lmn}(r)$, and it satisfies
\beq
\label{ring_commut}
\left[\hat\Psi_{lmn}(r),\hat\Pi_{l'm'n'}( r' )\right] = i\delta (r-r')\delta_{l,l'}\delta_{m,m'}\delta_{n,n'}
\eeq

As in the earlier calculations, to evaluate the  EE we need to discretize
the Hamiltonian. Here again the lattice spacing is $a$ and $R = a \,
Q$. Using the midpoint discretization scheme, the above Hamiltonian is
in the form of a system of $N$ coupled quantum harmonic oscillators
\ref{eq:discretizeHamiltonian} and can be written as a $N \times N$
symmetric semidefinite matrix (\ref{eq:Ring-Kij-Pert}).

The total entropy for the full Hamiltonian $H =\dis \sum_{l, m_{_1}, m_{_2}} H_{l m_{_1} m_{_2}}$, 
is given by
\bea
\label{eq:GS-ent2}
S (n',N)& =& \sum_{l, m,n} S_{l m n} (n',N)  \\
S_{l m n} (n',N) &=& - \ln[1-\xi_i] - \frac{\xi_i}{1 - \xi_i} \ln\xi_i \, .
\eea
where $\xi_i$ are given by Eq. (\ref{2HO-ent}).  

\begin{figure}[H]
\includegraphics[width=0.8\columnwidth]{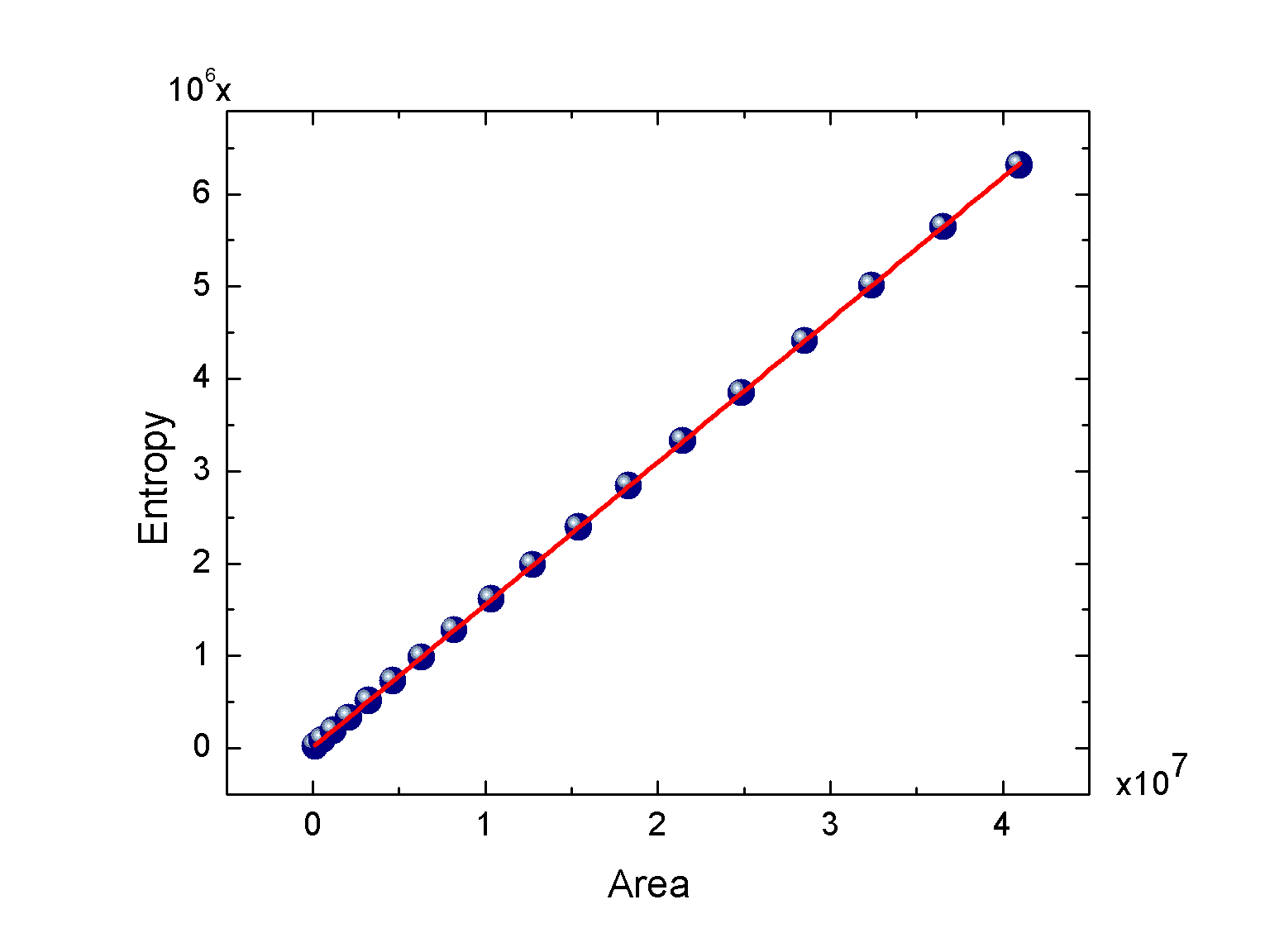} 
\caption{\footnotesize{Entropy versus scaled area profile in the $4+1$ D black rings using {\it perturbative} approach for $N=100,~ Q=5000, ~5\le n'\le 90$. 
The blue dots are the numerical outputs and the red line is best linear fit.}}
\label{Fig:Ring-Pert}
\end{figure}

In Fig. \ref{Fig:Ring-Pert}, we have plotted the EE
vs area of the $S^2 \times S^1$ surface.  As it is clear from the
figure, in the linear perturbative limit, the entropy is proportional
to area. We will discuss the importance of the result in
Sec. \ref{sec:Conclusion}.

\subsection{Constant angle approach}

By setting the angle $\phi_{_1}$ to a constant ($\alpha$), the action
(\ref{ring_action}) reduces to

\bea
\label{ring_const_angl_act}
S&=\dis\frac{1}{2}\sum_{m_{_1},m_{_2}}^\infty \int dt \thinspace dr \l[\l(\partial_t\widetilde{\Psi}_{m_{_1},m_{_2}}\r)^2-\l(1+\dis\frac{r}{R}\cos\alpha\r)^2\r.\nn\\
&\l.\dis \times\l[r(1-\frac{r^2}{R^2}) \l(\partial_r\l(\frac{\widetilde{\Psi}_{m_{_1},m_{_2}}}{\sqrt{r}}\r)+ \frac{3\cos\alpha}{2 R}\widetilde{\Psi}_{m_{_1},m_{_2}}\r)^2 \r.\r.\nn\\
&\l.\l.\dis +\l( \frac{m^2_{_1}}{r^2\sin^2\alpha}+\frac{m^2_{_2}}{R^2-r^2}\r)\widetilde{\Psi}^2_{m_{_1},m_{_2}}  \r]\r]
\eea
where $$\dis\widetilde{\Psi}_{m_{_1},m_{_2}}=\sqrt{\frac{R r \sin\alpha}{\l(1+\dis\frac{r}{R}\cos\alpha\r)}}\Phi_{m_{_1},m_{_2}}$$

The Hamiltonian corresponding to the above reduced action is:
\bea
H&=\dis\frac{1}{2}\sum_{m_{_1},m_{_2}}\int dr \l[\widetilde{\Pi}^2_{m_{_1},m_{_2}}+\l(1+\dis\frac{r}{R}\cos\alpha\r)^2\r.\nn\\
&\l.\dis \l[r(1-\frac{r^2}{R^2}) \l(\pa_r\l(\frac{\widetilde{\Psi}_{m_{_1},m_{_2}}}{\sqrt{r}}\r)+ \frac{3\cos\alpha}{2 R}\widetilde{\Psi}_{m_{_1},m_{_2}}\r)^2 \r.\r.\nn\\
&\l.\l.\dis +\l( \frac{m^2_{_1}}{r^2\sin^2\alpha}+\frac{m^2_{_2}}{R^2-r^2}\r)\widetilde{\Psi}^2_{m_{_1},m_{_2}}  \r]\r]
\eea
where $\widetilde{\Pi}_{m_{_1},m_{_2}}$ is canonically conjugate to
$\widetilde{\Psi}_{m_{_1},m_{_2}}$ and

$$\dis\widetilde{\Psi}_{m_{_1},m_{_2}}=\sqrt{\frac{R r
    \sin\alpha}{\l(1+\dis\frac{r}{R}\cos\alpha\r)}}\Psi_{m_{_1},m_{_2}}$$

\begin{figure}[H]
\includegraphics[width=1\columnwidth]{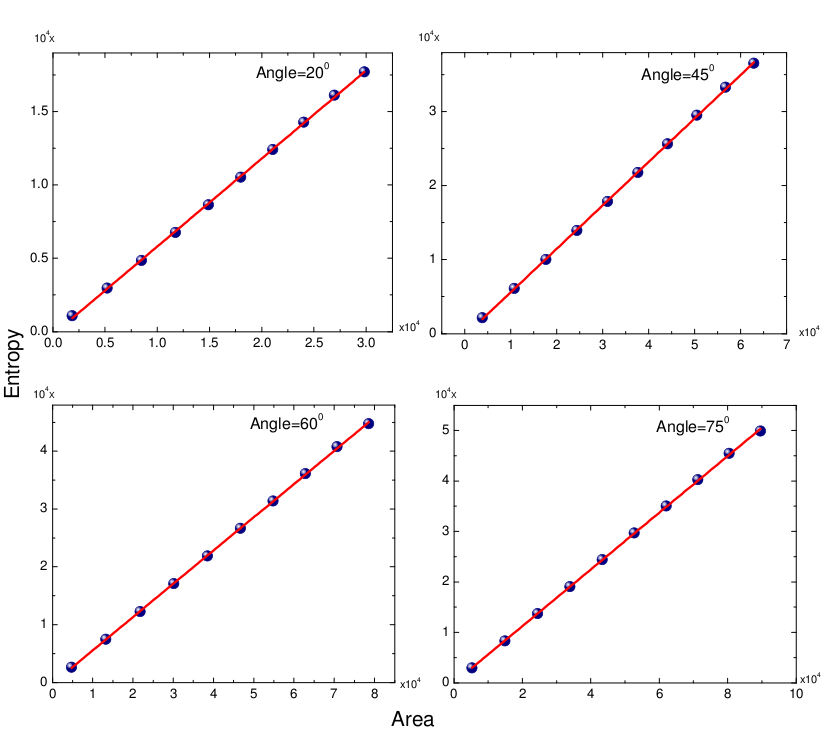} 
\caption{\footnotesize{Entropy versus scaled area profile in the $4+1$ D black rings using {\it constant angle} approach for$N=100,~ Q=5000, ~5\le n'\le 90$. 
The blue dots are the numerical outputs and the red line is best linear fit.}}
\label{meth3(4)}
\end{figure}

Discretizing the Hamiltonian and following the procedure discussed
above, we obtain the entropy for different angles.  Fig.
\ref{meth3(4)} shows the plot of entropy vs the area for
different angles. The plots show that the entanglement entropy is
linearly related to area.

%


\section{Conclusions and Discussions}
\label{sec:Conclusion}

In this work, we have obtained the entanglement entropy of massless,
minimally coupled scalar fields in Genus-1 topologies.  Specifically,
we have shown that the entanglement entropy is linearly related to the 
area of the torus ($S^1 \times S^1$) and ring geometry ($S^2 \times
S^1$).

Genus-1 topologies have lesser symmetry compared to Genus-0 topologies,
and hence we use {\it ab-initio} calculations to obtain entanglement
entropy in these topologies. One of the main difficulties in
evaluating entanglement entropy in Genus-1 topologies is that the
Helmholtz equation is not separable. This implies that the one will
not be able to write down the ground-state wave function exactly. To
circumvent this problem, we have used two complementary approaches to
evaluate entanglement entropy. In the first approach we have assumed
that the ratio of the smaller radius to the larger radius is much
less than unity. In the second approach, we evaluated the entanglement
entropy by setting one of the angular coordinates to be constant. In this
case, the entanglement entropy can be evaluated exactly as the
effective Helmholtz equation is separable.

Both these approaches clearly show that the entanglement entropy is
proportional to the area of the Genus-1 constant radius surface. In
several ways, the result cannot be extrapolated from the case of the 
sphere. First, the Genus-1 topologies are not simply connected
like the spheres. Second, it has been shown that  most of the
contribution to the entanglement entropy comes from close to the
surface \cite{shanki-sour}. In the case of Genus-1, since the surface
is not simply connected, it is not obvious that only the short-range
effects will dominate. Our analysis in this work, shows that this is
indeed the case.
  
The result brings attention to the following interesting questions.
Does the presence of mass to the scalar field affect the entanglement
entropy relation for Genus-1 topologies? Does the entanglement
entropy-area law hold for Genus-2 or higher surfaces? In Appendix B
we show that the constant angle approach for the spherical geometry
fixes the proportionality constant. It will be interesting to know
whether one can use this approach to analytically obtain entanglement
entropy in higher dimensions with the subleading corrections.

We hope to return to  study these problems in the near future.

\section*{Acknowledgments}
The work is supported by Max Planck-India Partner Group on Gravity and
Cosmology. S.S.K acknowledges CSIR, Government. of India, for the financial
support.  S.G acknowledges the hospitality at IISER-TVM, where the
initial part of this work was done.  S.S is partially supported by
Ramanujan Fellowship of DST, India. Part of the numerical computations
were performed on the supercomputing clusters at Albert Einstein
Institute, Golm.

\appendix

\section{Thin inner radii $(r/R \ll 1)$ approximation for Torus}
\label{app:approx1}

In this appendix, we calculate the entanglement entropy for a thin
torus, i. e., $r\ggg1$, such that

\beq
\Delta\sim \cosh r \sim \sinh r
\label{thin_tor_appr}
\eeq
This approximation (\ref{thin_tor_appr}) implies that $r$ takes a
minimum value, say $\beta$, which is always positive.  Substituting
Eq.(\ref{ansatz1}) in Eq. (\ref{tor_acti_form}) with this approximation,
we get
\bea
\label{thin_tor_actn1}
S&\sim\dis\frac{1}{2}\sum_{m_{_1},m_{_2}}\int dt \int_{\beta}^\infty q\thi dr\l[\frac{q^2}{\Delta^2}\l(\partial_t\Psi_{m_{_1},m_{_2}}\r)^2\r.\nn\\
&\l.-\dis  \thi\l(\partial_r\Psi_{m_{_1},m_{_2}}\r)^2
-\dis\left(m_{_1}^2+\frac{m_{_2}^2}{\sinh^2 r}\right)\Psi^2_{m_{_1},m_{_2}}\r]
\eea
Rewriting Eq.(\ref{thin_tor_actn1}) in terms of inner radius of the
torus, we get

\bea
\label{thin_tor_actn2}
S&=\dis\frac{1}{2}\sum_{m_{_1},m_{_2}}\int dt \int_0^\infty d\rho\l[\l(\partial_t\widetilde{\Psi}_{m_{_1},m_{_2}}\r)^2 \r.\nn\\
&\l. \dis-\rho\l[\partial_\rho\l(\frac{\dis\widetilde{\Psi}_{m_{_1},m_{_2}}}{\dis\sqrt{\rho}}\r)\r]^2
-\dis\l(\frac{m_{_1}^2}{\rho^2}+\frac{m_{_2}^2}{q^2 }\right)\widetilde{\Psi}^2_{m_{_1},m_{_2}}\r]
\eea
where  $ \displaystyle\sqrt{q \rho}\thinspace\Psi_{m_{_1},m_{_2}}= \widetilde{\Psi}_{m_{_1},m_{_2}}$.
The corresponding Hamiltonian is

 \bea
 \label{thin_tor_hamiltn1}
 H&=\dis\frac{1}{2}\sum_{m_{_1},m_{_2}}\int_0^\infty d\rho \thi \left[\thi \widetilde{\Pi}^2_{m_{_1},m_{_2}} +
  \rho\thi\left[ \partial_{\rho}\left(\frac{\widetilde{\Psi}_{m_{_1},m_{_2}}}{\sqrt{\rho}}\right)\right]^2\r.\nn\\
 &\l.\dis+\left(\frac{m_{_1}^2}{\rho^2}+\frac{m_{_2}^2}{q^2}\right)\widetilde{\Psi}^2_{m_{_1},m_{_2}}\right]
 \eea

The discretized Hamiltonian is  
\bea
\label{thin_tor_hamiltn2}
 H&=\dis\frac{1}{2 a}\sum_{m_{_1},m_{_2}}\sum_{j=1}^N \left[\widetilde{\Pi}^2_{m_{_1}m_{_2},j}
+\left(\frac{m_{_1}^2}{j^2}+\frac{m_{_2}^2}{Q^2}\right)\widetilde{\Psi}^2_{m_{_1}m_{_2},j}\r.\nonumber\\
&\l.\dis+(j+\frac{1}{2})\left(\frac{\widetilde{\Psi}_{m_{_1}m_{_2},j+1}}{\sqrt{j+1}}-\frac{\widetilde{\Psi}_{m_{_1}m_{_2},j}}{\sqrt{j}}\right)^2 \right] 
\eea

\begin{figure}[h]
\includegraphics[width=1\columnwidth]{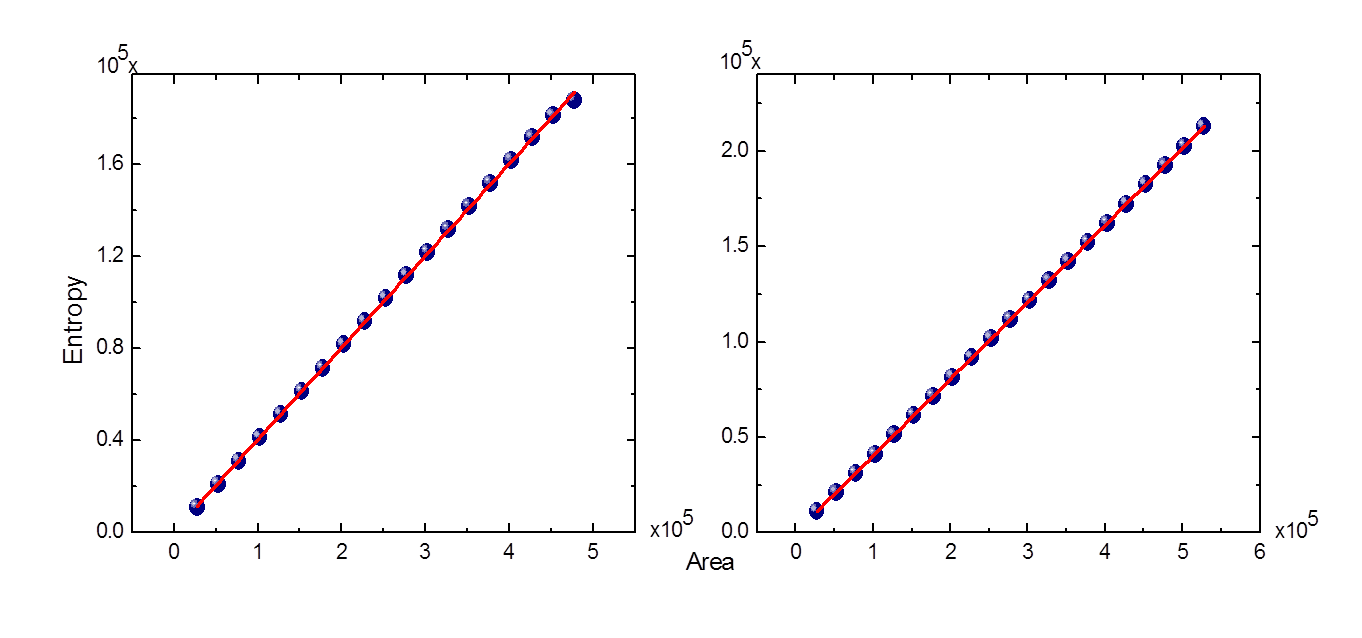} 
\caption{\footnotesize{Entropy versus scaled area profile in the $3+1$ D torus having thin inner radii with  $ N=100 (200), ~Q=5000,~ 5
 \le n' \le 95 \l(5 \le n' \le 195 \r) $ respectively. 
The blue dots are the numerical outputs and the red line is best linear fit.}}
\label{meth1(3)}
\end{figure}

In Fig. \ref{meth1(3)} we have plotted EE by following the procedure
discussed in Sec. II.  It is clear from the plot that the
entropy-area relation is satisfied in this limit.

\section{Constant angle- Spherical case}
\label{app:constantGenus0}

In this work, we have used the constant angle approach to obtain the
entanglement entropy for a Genus-1 surface.  The approach is
necessitated by the fact that in nonzero Genus topologies the
Helmholtz equation is not separable.

Applying this approach in the spherical case leads to an interesting
result. One of the criticisms of entanglement entropy is the that the
prefactor of the entanglement entropy-area relation is undetermined
\cite{wald}. In this appendix we show that the constant angle approach
may provide a plausible way of fixing the coefficient in the
entropy-area relation and that the approach may provide
subleading corrections to the area law.

\begin{figure}[H]
\includegraphics[width=1.0\columnwidth]{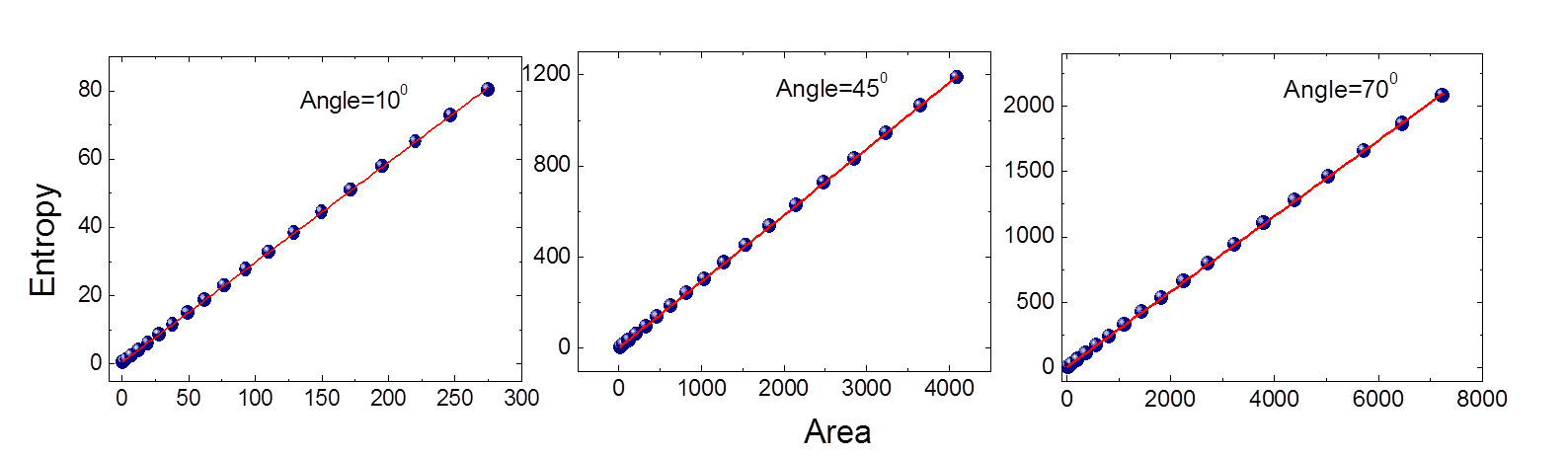} 
\caption{\footnotesize{Entropy versus reduced area for
    constant angle approach in the case of effective 3-D space
     with the number of sites is N=100($5\le n'\le 95$). The blue dots are
    the numerical outputs and the red line is best linear fit.}}
\label{Fig:Effective3D}
\end{figure}

Recently, one of the authors along with Das and Braunstein
\cite{Braunstein:2011sx} have shown that the entanglement entropy-area
law is valid for $D$-Sphere $D > 1$. 
To understand the importance of the constant angle approach, 
let us set the polar angle ($\theta=\alpha$) to be a constant value in the 
3-Sphere ($ r, \theta, \phi_{_1},\phi_{_2} $). The reduced line element is,
\beq
\label{eq:effective_3d_line}
ds^2= dt^2-\l[dr^2+ r^2 \sin^2\alpha\l(d\phi^2_{_1}+\sin^2\phi_{_1} d\phi^2_{_2} \r)\r]
\eeq 
The above equation is similar to an effective line element in the $3+1$ -dimensional for
which the Hamiltonian is
\beq
\label{ eq:effective_3d_hamiltonian}
H=\frac{1}{2} \sum_{l,m} \int dr \l[\widetilde{\Pi}^2_{_{lm}}+r^2 \l[\pa_r\l(\frac{\widetilde{\Psi}_{{lm}}}{r}\r)\r]^2
+\frac{l(l+1)}{r^2 \sin^2\alpha}\widetilde{\Psi}^2_{_{lm}} \r]
\eeq  
where we have used the  following ansatz for expanding the action 
$$\Phi(t,r,\phi_{_1},\phi_{_2})= \sum_{l,m}\frac{\widetilde{\Psi}_{_{lm}}(t,r)}{r\sin\alpha} \mathcal{Z}_{_{l,m}}(\phi_{_1},\phi_{_2}) \, .$$
From the Fig. \ref{Fig:Effective3D}, it is interesting to note that the EE varies linearly with  the angle-dependent area 
(proportional to $\sin^2\alpha$), which is identical to the case of the three -dimensional sphere \cite{1993-Srednicki-PRL} with the slope being $0.29$.

In the same way, one can start with the line element in the $(3+1)$ -dimensional space-time ($t,r,\theta,\phi$) and set the polar angle to 
be a constant leading to an effective line element in the $(2 + 1)$ -dimensional space-time
\beq
\label{eq:effective_2d_line}
ds^2= dt^2-\l[dr^2+ r^2 \sin^2\alpha d\phi^2\r]
\eeq 
The form of the effective two -dimensional Hamiltonian is,
\beq
\label{ eq:effective_2d_hamiltonian}
H=\frac{1}{2} \sum_{m} \int dr \l[\widetilde{\Pi}^2_{_m}+r \l[\pa_r\l(\frac{\widetilde{\Psi}_{m}}{\sqrt{r}}\r)\r]^2
+\frac{m^2}{r^2 \sin^2\alpha}\widetilde{\Psi}^2_{_m} \r]
\eeq  
Again the ansatz has  the form,
$$ \Phi(t,r,\phi)= \sum_{m=-\infty}^{\infty}\frac{\widetilde{\Psi}_{_m}(t,r)}{\sqrt{\pi r\sin\alpha}}\cos m\phi$$

\begin{figure}[H]
\includegraphics[width=1.0\columnwidth]{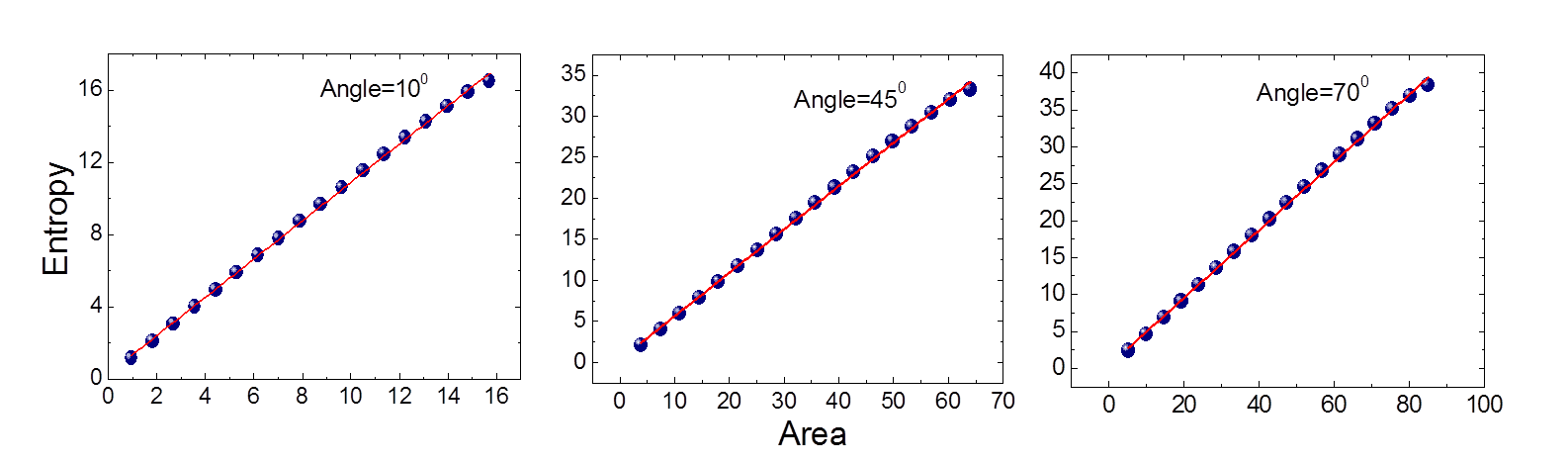} 
\caption{\footnotesize{Entropy versus the reduced area for
    constant angle in the case of effective 2-D space
     with the number of sites is N=100($5\le n'\le 95$). The blue dots are
    the numerical outputs and the red line is best linear fit.}}
\label{Fig:Effective2D}
\end{figure}

\noindent Repeating the same procedure as discussed earlier, the plot of the EE as a function of area showing that the entropy is proportional to 
$sin \alpha$ is shown in Fig\ref{Fig:Effective2D}.

In the same way, the entanglement entropy for the scalar fields in the reduced $(2 + 1)$ -dimensional space-time leads to the  
relation
\beq
\label{entropy1d}
S_{ent}^{(1+1)}= k_{_0} \ln \left(r/a \right)
\eeq
where $a$ is the lattice spacing and $k_{_0}$ is a constant \cite{Cardy-Calabrese-2004}. Unlike 
the higher-dimensional space-times, in this case the constant angle entanglement entropy is 
{\it independent} of the angle. This provides an interesting possibility of obtaining the higher 
-dimensional entanglement entropy from the above logarithmic dependence (\ref{entropy1d}). 
Specifically, it is interesting to note that the entanglement entropy for the scalar field 
in $(2+1)$-dimensional space-time can be obtained by integrating Eq. (\ref{entropy1d}) with respect to
$\alpha$, i. e.,

\br
\label{entropy2d}
S_{ent}^{(2+1)}&=& \int_{\alpha=0}^{\alpha=2\pi} r d\alpha \thi k_{_0}
\ln(r/a)\nn\\ &=&2\pi\thi r k_{_0} \ln (1/a) + 2\pi\thi r k_{_0 }\ln
r\nn\\ &=&2\pi r k_{_1}+ 2\pi\thi r k_{_0} \ln r 
\er 
where $k_{_1}=k_{_0} \ln (1/a)$. Note that $a \ll 1$; hence $k_{_1} > 0.$
A similar procedure can be applied to obtain the EE for $3+1$ dimensions:
\beq
\label{entropy3d}
 S_{ent}^{(3+1)}=4\pi r^2k_{_1}+ k_{_0} r^2 \int_{\alpha=0}^{\alpha=\pi} d\alpha \sin\alpha \ln (r^2\sin\alpha)
\eeq
Thus, by fixing the constant $k_{_0}$ we can calculate the corrections to the
EE for $(D+1)$-sphere symmetric space-time as
\beq
\label{entropy4d}
 S_{ent}^{(D+1)}=4\pi r^{D-1}k_{_1}+ f(k_{_0})
\eeq
where $f(k_{_0})$ is the correction to the $D+1$ -dimensional entropy,
which is a function of $k_{_0}$ only. It interesting to note that the area-
dependent terms only depend on $k_{_1}$ and are determined. we have
 yet to understand the full implications of the above result and it
is currently being investigated.



\begin{widetext}

\section{Matrix elements in toroidal and ring geometries}

The interaction matrix that is used for computing the EE in all different
methods is given. The Hamiltonian of the system can be written
as,

\beq
\label{eq:discretizeHamiltonian}
\dis H=\frac{1}{2} \sum_{i=1}^N \Pi^2(x) + \frac{1}{2}\sum_{ i,j=1 }^N x_i K_{ij} x_{j} \\
\eeq 
where $K_{ij}$ is the interaction matrix elements.

\begin{itemize}
\item{\it Perturbative approach in toroidal co-ordinates}
\bea
K_{ij}& =\dis \l[ \frac{3}{2}-\frac{1}{Q}+\l(m^2_{_1}+\frac{m^2_{_2}}{Q^2}\r)\l( 1-\frac{2}{Q}\r)\r]
\delta_{i1}\delta_{j1} +\dis\l[ 2+\frac{1}{Q}\l( 1-4j+\frac{1}{j}\r)
\dis +m^2_{_1}\l(\frac{1}{j^2}-\frac{2}{Qj}\r)+\frac{m^2_{_2}}{Q^2}\l(1-\frac{2j}{Q}\r) \r]\delta_{ij}\nn\\
\label{eq:Torus-Kij-Pert}
& \dis + \l[\frac{2(j+\frac{1}{2})}{Q}-\frac{1}{2Q}- \frac{j+\frac{1}{2}}{\sqrt{j(j+1)}}\r]\delta_{i j+1}
\dis+ \l[\frac{2(i+\frac{1}{2})}{Q}-\frac{1}{2Q}- \frac{i+\frac{1}{2}}{\sqrt{i(i+1)}}\r]\delta_{i j-1}
\eea
 \item {\it Constant angle approach in toroidal co-ordinates}
\bea
K_{ij}& =\dis \l[\frac{\l(\sqrt{1+Q^2}-\cos\alpha\r)^2}{Q^4}\l(\dis \sqrt{\l(\frac{9}{4}+Q^2\r)\l(1+Q^2\r)}+m^2\r)\r]\delta_{i1}\delta_{j1}
+\l[\dis\frac{\l(\sqrt{1+\dis\l(\frac{Q}{j}\r)^2}-\cos\alpha\r)^2}{Q^4}j^2\sqrt{j^2+Q^2}\r.\nn\\
&\l. \dis \times\l(\sqrt{\dis\l(j-\frac{1}{2}\r)^2+Q^2}+\sqrt{\dis\l(j+\frac{1}{2}\r)^2+Q^2}+\dis\frac{m^2}{\sqrt{j^2+Q^2}} \r) \r]\delta_{ij}
\dis -\l[\frac{\sqrt{\l(j+\frac{1}{2}\r)^2+Q^2}}{Q^4}\r.\nn\\
&\l.\dis\times j (j+1)\l(((j+1)^2+Q^2)(j^2+Q^2)\r)^{1/4}
\dis\l(\sqrt{1+\dis\l(\frac{Q}{j}\r)^2}-\cos\alpha\r)\dis\l(\sqrt{1+\dis\l(\frac{Q}{j+1}\r)^2}-\cos\alpha\r)\r]\delta_{i j+1}\nn\\
& \dis -\l[\frac{\sqrt{\l(i+\frac{1}{2}\r)^2+Q^2}}{Q^4}i (i+1)\l(((i+1)^2+Q^2)(i^2+Q^2)\r)^{1/4}
\dis\l(\sqrt{1+\dis\l(\frac{Q}{i}\r)^2}-\cos\alpha\r)\r.\nn\\
\label{eq:Torus-Kij-ConstAngle}
&\l.\dis\times\l(\sqrt{1+\dis\l(\frac{Q}{i+1}\r)^2}-\cos\alpha\r)\r]\delta_{i j-1}
\eea

\item {\it Perturbative approach in ring co-ordinates}
\bea
K_{ij}& =\dis \l[ \frac{9}{4}+l(l+1)+\frac{n^2}{Q^2}+C_0\l(\frac{4n^2}{Q^3}+\frac{4l(l+1)}{Q}-\frac{21}{4Q}\r)\r]\delta_{i1}\delta_{j1}
+\l[ 2+\frac{1}{2j^2}+\frac{l(l+1)}{j^2}+\frac{n^2}{Q^2}\r.\nn\\
&\l. \dis +C_0\l( \frac{4j}{Q}+\frac{3}{Qj}-8\frac{j+\frac{1}{2}}{jQ}+4\frac{n^2j}{Q^3}+4\frac{l(l+1)}{jQ}\r)\r]\delta_{ij}
+ \dis\l[-\frac{\l(j+\frac{1}{2}\r)^2}{j(j+1)} +\frac{2C_0\dis\l(j+\frac{1}{2}\r)}{Q(j+1)} \l(2-\frac{\l(j+\frac{1}{2}\r)^2}{j}\r)\r]\delta_{i j+1}\nn\\
\label{eq:Ring-Kij-Pert}
&+ \dis\l[-\frac{\l(i+\frac{1}{2}\r)^2}{i(i+1)} +\frac{2C_0\dis\l(i+\frac{1}{2}\r)}{Q(i+1)} \l(2-\frac{\l(i+\frac{1}{2}\r)^2}{i}\r)\r]\delta_{i j-1}
\eea
\item {\it Constant angle approach in ring co-ordinates}
\bea
K_{ij}& =\dis \l[\frac{\dis\frac{3}{2}\l(1-\frac{9}{4Q^2}\r)}{1+\dis\frac{3}{2Q}\cos\alpha}\l(1+\dis\frac{\cos\alpha}{Q}\r)^3
 +\dis \l(\frac{m^2_{_1}}{\sin^2\alpha}+\frac{m^2_{_2}}{Q^2-1}\r)\l(1+\dis\frac{\cos\alpha}{Q}\r)^2\r]\delta_{i1}\delta_{j1}+ \l(1+\dis\frac{j\cos\alpha}{Q}\r)^3\nn\\
&\times\l[\dis \frac{\l(j-\frac{1}{2}\r) \l(1-\dis \frac{(j-\frac{1}{2})^2}{Q^2}\r)}{j\l(1+\dis\frac{(j-\frac{1}{2})\cos\alpha}{Q}\r)} 
+ \dis \frac{\l(j+\frac{1}{2}\r) \l(1-\dis \frac{(j+\frac{1}{2})^2}{Q^2}\r)}{j\l(1+\dis\frac{(j+\frac{1}{2})\cos\alpha}{Q}\r)}
+\dis \l(\frac{m^2_{_1}}{j^2\sin^2\alpha}+\frac{m^2_{_2}}{Q^2-j^2}\r)\l(1+\dis\frac{j\cos\alpha}{Q}\r)^{-1}\r]\delta_{ij}\nn\\
& -\l[\dis\frac{\l(j+\frac{1}{2}\r) \l(1-\dis \frac{(j+\frac{1}{2})^2}{Q^2}\r)}{\l(1+\dis\frac{(j+\frac{1}{2})\cos\alpha}{Q}\r)} 
\frac{\dis\l(1+\frac{j\cos\alpha}{Q}\r)^{\dis\frac{3}{2}}}{\sqrt{j(j+1)}}\dis\l(1+\frac{(j+1)\cos\alpha}{Q}\r)^{\dis\frac{3}{2}} \r] \delta_{i j+1}\nn\\
\label{eq:Ring-Kij-ConstAngle}
& -\l[\dis\frac{\l(i+\frac{1}{2}\r) \l(1-\dis \frac{(i+\frac{1}{2})^2}{Q^2}\r)}{\l(1+\dis\frac{(i+\frac{1}{2})\cos\alpha}{Q}\r)} 
\frac{\dis\l(1+\frac{i\cos\alpha}{Q}\r)^{\dis\frac{3}{2}}}{\sqrt{i(i+1)}}\dis\l(1+\frac{(i+1)\cos\alpha}{Q}\r)^{\dis\frac{3}{2}} \r] \delta_{i j-1}
\eea

\end{itemize}
\end{widetext}
\section{Evaluation of integrals }
The integrals in Eq.(\ref{ring_action}) are listed below  and can be  evaluated exactly \cite{arfken,Rashid}:
\begin{subequations}
 \label{pert_integrals1}
\br
I_{_1}&=&\int_0^\pi\int_0^{2\pi} \mathcal{Z}_{lm} \cos\theta \mathcal{Z}_{l'm'} d\Omega  \\
I_{_2}&=&\int_0^\pi\int_0^{2\pi} \partial_{\theta}\mathcal{Z}_{lm} \cos\theta \partial_{\theta}\mathcal{Z}_{l'm'} d\Omega\\
I_{_3}&=&\int_0^\pi\int_0^{2\pi} \mathcal{Z}_{lm}\sin\theta\partial_{\theta}\mathcal{Z}_{l'm'} d\Omega\\
I_{_4}&=&\int_0^\pi\int_0^{2\pi}  \partial_{\phi_{_1}}\mathcal{Z}_{lm} \frac{\cos\theta}{\sin^2\theta} \partial_{\phi_{_1}}\mathcal{Z}_{l'm'} d\Omega
\er
\end{subequations}

\begin{subequations}
 \label{pert_integrals2}
\br
I_{_1}&=& \l[C_0 \delta_{l',l-1}+C_1\delta_{l',l+1}\r]\delta_{m,m'}\\
I_{_3}&=&\l[-(l+1)C_0 \delta_{l',l-1}+l C_1 \delta_{l',l+1}\r]\delta_{m,m'}\\
I_{_4}&=&\frac{m}{2}\l[(2l-1)C_0 \delta_{l',l-1}\r.\nn\\
&&\l.\dis\qquad\qquad +(2l+3) C_1 \delta_{l',l+1}\r]\delta_{m,m'}\\
I_{_2}&=& l(l+1) I_{_1} +I_{_3}-I_{_4}
\er
\end{subequations}

where, $ d\Omega= \sin\theta d\theta d\phi_{_1}$ and 
\\$ C_0= \dis\sqrt{\frac{l^2-m^2}{4l^2-1}},\;\;\; C_1= \dis \sqrt{\frac{\l(l-m+1\r)\l(l-m+1\r)}{\l(2l+1\r)\l(2l+3\r)}}$.

\vskip 0.16truein

\end{document}